\documentclass[a4paper,10pt]{article}
\usepackage[utf8]{inputenc}
\usepackage[fleqn]{amsmath}
\usepackage{amsfonts}
\usepackage{amssymb} 
\title{ Boundary Effects in Super-Yang-Mills Theory}
\author{Mushtaq B Shah$^1$, Mir Faizal$^{2, 3}$, Prince A Ganai$^1$,\\ Zaid Zaz$^4$,  Anha Bhat$^5$, Syed Masood$^6$,\\\\
\\$^1$Department of Physics, National Institute of Technology,\\
Srinagar, Kashmir-190006, India \\ 
$^2$Irving K. Barber School of Arts and Sciences,
\\ University of British
Columbia - Okanagan \\
  Kelowna,  British Columbia V1V 1V7, Canada
\\$^3$Department of Physics and Astronomy,
University of Lethbridge,  \\ 
Alberta, T1K 3M4, Canada \\ $^4$Department of Electronics and Communication Engineering , \\University of Kashmir, 
Srinagar, Kashmir-190006, India 
\\$^5$  Department of Metallurgical and Materials Engineering, \\ National Institute of Technology, \\
Srinagar, Kashmir-190006, India\\ 
$^6$Department of Physics,
International Islamic University,\\ 
H-10 Sector, Islamabad, Pakistan}
\date{}
\begin{document}

\maketitle

\begin{abstract}
In this paper, we shall analyse a three dimensional supersymmetry theory  
with  $\mathcal{N} = 2$.  
The effective Lagrangian  will  be given by 
the sum of the gauge fixing term and the ghost term with the original  classical  
Lagrangian. In presence of a boundary the supersymmetry of this Lagrangian will be broken. 
However, it will be possible to preserve half the supersymmetry    even in presence of a boundary. 
This will be done by adding a boundary Lagrangian to the effective bulk Lagrangian. 
The supersymmetric transformation of this new boundary Lagrangian will exactly cancel the boundary term generated 
from the supersymmetric transformation of the effective bulk Lagrangian.  We will obtain the  Slavnov-Taylor Identity  
for this theory. 
\end{abstract}
 
\section{Introduction}
As any gauge theory contains unphysical gauge degrees of freedom, and it is not possible quantize this theory 
without removing these unphysical degrees of freedom. This is achieved by fixing a gauge, and the gauge fixing 
is incorporated at a quantum level by adding gauge fixing term to the original Lagrangian. We also need 
to add a ghost term corresponding to this gauge fixing term to the original Lagrangian. 
 This new effective Lagrangian obtained from a sum of the  
  original 
classical Lagrangian with the gauge fixing and the ghost terms
is  invariant under the BRST transformations \cite{brst}-\cite{brst1}. The BRST symmetry has been  studied  
for various different gauges \cite{nlbrst}-\cite{nlbrst1}, and  
has been applied for analysing various aspects of different 
supersymmetric theories  \cite{mir}-\cite{sudd}.

The BRST symmetry has also been used in analysing 
   ghost-anti-ghost condensation  \cite{kon}-\cite{z4}.   
Furthermore, such  ghost-anti-ghost condensation has been proposed 
as the mass providing mechanism  of the off-diagonal gluons and off-diagonal ghosts
in the Yang-Mills theory   \cite{sch}-\cite{kon1}. This analysis has been performed using the 
the Maximal Abelian gauge. 
An evidence for infrared Abelian dominance has also been 
provided by this mechanism \cite{hooft}, thereby justifying the dual
superconductor picture \cite{nambu}-\cite{poly} of QCD vacuum. This has been used 
in explaining quark confinement \cite{kon1}-\cite{kon4}. 
It may be noted that interesting consequences of the breaking of BRST symmetry 
have also been discussed \cite{kon}-\cite{n4}.

The action for most 
 renormalizable quantum field theories,  including supersymmetric theories, 
is at most
quadratic in derivatives. So, the supersymmetric variation of such an action   produces   
a total derivative term. In absence of a boundary this 
  total derivative term vanishes. 
 However, in presence of a boundary,
boundary contributions arise due to such a total derivative term. This breaks the supersymmetry of a supersymmetric 
theory in presence of a boundary. It may be noted that the translational invariance of any theory 
is broken by the presence of a boundary. The breaking of the translational invariance in a supersymmetric theory 
also breaks the supersymmetry of that theory.  However, it is possible to retain some on-shell
supersymmetry by imposing suitable 
 boundary conditions   \cite{a}-\cite{b}. The 
supersymmetry of a theory 
generates various constraints on the possible boundary conditions  \cite{c}-\cite{g}.

Even though some on-shell supersymmetry can be retained by imposing boundary conditions, the 
off-shell supersymmetry is still broken. This is because these boundary conditions are only imposed on the on-shell
field. It is important to preserve the off-shell supersymmetry  of a theory. This is because the 
 path integral formalism   uses off-shell fields, 
 and most   supersymmetric theories are quantized using a path integral formalism. 
 So, it is important  to preserve the off-shell supersymmetry for a theory.  
It is possible to preserve half the off-shell supersymmetry of a theory by modifying the original action 
of the theory.  
This has been done by the addition of new boundary terms to the original bulk action.
The boundary contribution generated from the supersymmetric variation of the original bulk action
are exactly canceled by the supersymmetric variation of these  new boundary terms. 
This has been studied for a three dimensional theory with 
$\mathcal{N} =1 $ supersymmetry   \cite{1}. Furthermore, this procedure has been used for analysing 
 Chern-Simons-matter theories in presence of a boundary  \cite{4}-\cite{6}.
 It may be noted that  
 an additional  boundary term is also generated from the gauge transformation
 of Chern-Simons-matter theories in presence of a boundary. So, additional boundary degrees of freedom are needed to 
 preserve the gauge invariance of  a Chern-Simons-matter  theory  in presence of a boundary. 
This is because the boundary   contribution generated from the gauge
transformation of
the bulk action are exactly canceled by 
the gauge transformation of   these new boundary degrees of freedom.
 
Non-anticommutative deformation of   supersymmetric theories   
has also been studied using this off-shell formalism \cite{8}. 
This has been done for a theory with $\mathcal{N} =2$ supersymmetric theory in three dimensions. 
In this analysis, half the supersymmetry of such a
supersymmetric theory was broken by imposing non-anticommutativity.  Then by  
  suitably combining the boundary effects with   
non-anticommutativity, a theory with $\mathcal{N} = 1/2$ supersymmetry 
was constructed. A three dimensional 
super-Yang-Mills  theory  has also been coupled 
to background flux in presence of a boundary \cite{9}. 
In this paper,we shall do an analysis of gauge theory with  
with $\mathcal{N} =2$ supersymmetry in presence of a boundary. It is important to perform such an analysis to demonstrate the preservation 
of half the supersymmetry for a gauge theory at a quantum level. So, we will analyze this theory using the quantum fluctuations around 
a fixed background. We will analyze the BRST symmetry of such a theory, by analyzing the fields as a sum of the classical background 
fields and quantum fluctuations around such classical fields. We will also analyze the Slavnov-Taylor Identity  for such a theory.

\section{Supersymmetric Gauge  Theory}
In this section, we will review the construction of a three dimensional  
supersymmetric gauge theory in presence of a boundary \cite{1}-\cite{8}. 
We define two fermionic coordinates, $\theta_{1a} = (\theta_{11}, \theta_{12})$ and $\theta_{2a} = (\theta_{2a}, \theta_{22} )$. 
Now we can also define $(\gamma^\mu \theta_1)_a = (\gamma^\mu)^b_a \theta_{1a}$ and $(\gamma^\mu \theta_2)_a = (\gamma^\mu)^b_a \theta_{2a}$. 
The raising and lowering of the spinor indices occurs as $\theta^a_1 = C^{ba}\theta_{1 b}, \theta_{1a} = \theta_1 ^b C_{ab} $, and 
$\theta^a_2 = C^{ba}\theta_{2 b}, \theta_{2a} = \theta_2 ^b C_{ab} $. Here 
$C^{ab}= - C^{ba}, C_{ab} = - C_{ba}$, and $C^{ba}C_{cb}= \delta^a_c$. We also have $(\gamma^\mu)_{ab} = (\gamma^\mu)^c_a C_{cb}  = (\gamma^\mu)_{ba}$.
A  $\mathcal{N} = 2 $ supersymmetric theory in three dimensions can be parameterized by
   two supercharges,
\begin{eqnarray}
Q_{1a}  = \partial_{1a}  - (\gamma^\mu \theta_1)_a \partial_\mu, &&
Q_{2a}  = \partial_{2a}  - (\gamma^\mu \theta_2)_a \partial_\mu.
\end{eqnarray}
  These supercharges satisfy, \begin{eqnarray}
 \{Q_{1a}, Q_{1b}\} = 2 \gamma_{ab}^{\mu}\partial_\mu, && 
  \{Q_{2a}, Q_{2b}\} = 2 \gamma_{ab}^{\mu}\partial_\mu,  \nonumber \\ 
  \{Q_{1a}, Q_{2b}\} = 0. &&  \end{eqnarray}
Now we  define superderivatives as
\begin{eqnarray}
  D_{1a} = \partial_{1a} + (\gamma^\mu \theta_1)_a \partial_\mu,&& 
  D_{2a} = \partial_{2a} + (\gamma^\mu \theta_2)_a \partial_\mu. \end{eqnarray}
These superderivatives commute with the generators of 
  $\mathcal{N} = 2$ supersymmetry, $\{Q_{1a}, D_{1b}\} = \{Q_{1a}, D_{2b}\} =0 $ and $
 \{Q_{2a}, D_{1b}\}= \{Q_{2a}, D_{2b}\} =0. $ 
  These superderivatives also satisfy, \begin{eqnarray}
 \{D_{1a}, D_{1b}\} = -2 \gamma_{ab}^{\mu}\partial_\mu,  &&
  \{D_{2a}, D_{2b}\} =- 2 \gamma_{ab}^{\mu}\partial_\mu,  \nonumber \\ 
  \{D_{1a}, D_{2b}\} = 0. && \end{eqnarray}

We can also define  
gauge valued   spinor superfields $\Gamma_{1a}= \Gamma_{1a}^A  (\theta_1) T_A$ and $\Gamma_{2a}
 = \Gamma_{2a}^A  (\theta_2) T_A $,  where $[T_A, T_B] = i f^C_{AB} T_C$.  
 Now we can define covariant derivatives with these fields as 
 \begin{eqnarray}
  \nabla_{1a} = D_a -i \Gamma_{1a}, &&  \nabla_{2a} = D_a -i \Gamma_{2a}. 
 \end{eqnarray}
These fields transform   under the
  gauge transformation as ,
$
\Gamma_{1a}  \to i  u\nabla_{1a} u^{-1},$  and $
 \Gamma_{2a}  \to i u\nabla_{2a} u^{-1} 
$ \cite{gate}. 
We can also construct the field strengths as follows,
\begin{eqnarray}
W_{1a} &=&  \frac{1}{2} D^b_1 D_{1a} \Gamma_{1b} - \frac{i}{2}  \{\Gamma^b_1, D_{1b} \Gamma_{1a}\}
- \frac{1}{6} [ \Gamma^b_1 ,
\{ \Gamma_{1b}, \Gamma_{1a}\}],\nonumber \end{eqnarray}
\begin{eqnarray}
W_{2a} &=&  \frac{1}{2} D^b_2 D_{2a} \Gamma_{2b} - \frac{i}{2}  \{\Gamma^b_2, D_{2b} \Gamma_{2a}\}
- \frac{1}{6} [ \Gamma^2_1 ,
\{ \Gamma_{2b}, \Gamma_{2a}\}].
\end{eqnarray}
These field strengths transform as $
W_{1a}  \to  u W_{1a} u^{-1},$  and $
W_{2a}  \to  u W_{2a} u^{-1}.
$
We can write the action for super-Yang-Mills theory as
\begin{eqnarray}
 \mathcal{L} &=&  
  D^2_1 [W^a_1W_{1a}]_{\theta_1 =0} + D^2_2
[W^a_2W_{2a}]_{\theta_2 =0}. 
\end{eqnarray}
 
  In the presence of a boundary the supersymmetry is broken. However, half of the supersymmetry of the 
  original theory can be preserved by either adding or subtracting a boundary term to the original Lagrangian 
  \cite{1}. We now define a boundary along $x_3$ direction. Thus, we can define the boundary fields as fields 
  restricted to the boundary, and we can also construct boundary Lagrangian from such fields. 
We can define    $\mathcal{L}_{1b}$ and $\mathcal{L}_{2b}$ to be such  boundary Lagrangian constructed from 
the boundary fields. Now these boundary Lagrangian can be 
added or 
 subtracted from the bulk Lagrangian with $\mathcal{N} =2$ supersymmetry. 
 It is possible to  choose these boundary Lagrangian, such that 
  $ \mathcal{L} \pm \mathcal{L}_{1b}$    preserves the
supersymmetry generated by
$\epsilon^{1\mp} Q_{1\pm}$,
and $\mathcal{L} \pm \mathcal{L}_{2b}$    preserves the
supersymmetry generated by
$\epsilon^{2\mp} Q_{2\pm}$ \cite{8}.
Here the projection operators    $P_{\pm} = (1 \pm \gamma^3)/2$ have been used to obtain these projections of the supercharges. 
Now as the 
  original Lagrangian $\mathcal{L} = D_1^2    [\Omega_1(\theta_1)]_{\theta_1 = 0 }$ and $  \mathcal{L} = 
D_2^2    [\Omega_2 (\theta_2)]_{\theta_2 = 0 }$, 
  the boundary  terms  can be written as  $\mathcal{L}_{1b} =   \partial_3    [\Omega_1(\theta_1)]_{\theta_1   =0}  $ and $ 
\mathcal{L}_{2b} = \partial_3 [\Omega_2 (\theta_2)]_{ \theta_2 =0}$ \cite{4}. 
It is not possible to simultaneously preserve both the supersymmetry generated by  $\epsilon^{1-} Q_{1+}$ and
$\epsilon^{1+} Q_{1-}$, or  $\epsilon^{2-} Q_{2+}$ and
$\epsilon^{2+} Q_{2-}$,
in the presence of a boundary. 
However, in the presence of a boundary, we can construct the Lagrangian
which preserves the supersymmetry
generated by $\epsilon^{1\mp} Q_{1\pm}$ and $\epsilon^{2\mp} Q_{2\pm}$. 
We can write the Lagrangian for super-Yang-Mills theory which preserves various supersymmetries as  
 \cite{8},
\begin{eqnarray}
 \mathcal{L}^{1- 2 -} &=&  (D^2_1 - \partial_3) [W^a_1W_{1a}]_{\theta_1 =0} + (D^2_2 - \partial_3) [ W^a_2W_{2a}]_{\theta_2 =0}
 , \nonumber \end{eqnarray}
\begin{eqnarray}
 \mathcal{L}^{1- 2 +} &=& (D^2_1 - \partial_3) [W^a_1W_{1a}]_{\theta_1 =0} + (D^2_2+ \partial_3)[ W^a_2W_{2a}]_{\theta_2 =0}
 , \nonumber \end{eqnarray}
\begin{eqnarray}
 \mathcal{L}^{1+ 2 -} &=& (D^2_1 + \partial_3) [W^a_1W_{1a}]_{\theta_1 =0} + (D^2_2 - \partial_3)[ W^a_2W_{2a}]_{\theta_2 =0}
 , \nonumber \end{eqnarray}
\begin{eqnarray}
 \mathcal{L}^{1+ 2 + } &=&     (D^2_1 + \partial_3) [W^a_1W_{1a}]_{\theta_1 =0} + (D^2_2 + \partial_3) [ W^a_2W_{2a}]_{\theta_2 =0}.
\end{eqnarray}

 \section{BRST Symmetry} 
 In this section, we will study the effective Lagrangian obtained by the sum of the gauge fixing term and the ghost 
 term with the modified super-Yang-Mills Lagrangian in the Lorentz gauge. The  
  Lorentz gauge fixing conditions can be incorporated in the modified super-Yang-Mills Lagrangian
 at a quantum level by adding the following gauge fixing term to it 
 \begin{eqnarray}
   \mathcal{L}^{1+ 2+}_{gf} &=&    (D^2_1 + \partial_3) [ b_1   (D^a_1 \Gamma_{1a}) + \frac{\alpha}{2}b^2_1 ]_{\theta_1 =0} \nonumber \\
   && + (D^2_2 + \partial_3) 
   [  b_2  (D^a_2 \Gamma_{2a}) + \frac{\alpha}{2}b^2 ]_{\theta_2 =0}
 ,
\nonumber \end{eqnarray}
\begin{eqnarray}
   \mathcal{L}^{1- 2-}_{gf} &=&    (D^2_1 - \partial_3) [ b_1   (D^a_1 \Gamma_{1a}) + \frac{\alpha}{2}b^2_1]_{\theta_1 =0} 
   \nonumber \\ &&  + (D^2_2 - \partial_3) [  b_2  (D^a_2 \Gamma_{2a}) + \frac{\alpha}{2}b^2]_{\theta_2 =0}
 ,
\nonumber \end{eqnarray}
\begin{eqnarray}
   \mathcal{L}^{1+ 2-}_{gf} &=&  
   (D^2_1 + \partial_3) [ b_1   (D^a_1 \Gamma_{1a}) + \frac{\alpha}{2}b^2_1]_{\theta_1 =0}  \nonumber \\
   && + (D^2_2 - \partial_3) [ b_2  (D^a_2 \Gamma_{2a}) + \frac{\alpha}{2}b^2]_{\theta_2 =0}
 ,
\nonumber \end{eqnarray}
\begin{eqnarray}
   \mathcal{L}^{1- 2+}_{gf} &=&  
  (D^2_1- \partial_3) [ b_1   (D^a_1 \Gamma_{1a}) + \frac{\alpha}{2}b^2_1]_{\theta_1 =0}  \nonumber \\ &&  
  + (D^2_2 + \partial_3) [  b_2  (D^a_2 \Gamma_{2a}) + \frac{\alpha}{2}b^2]_{\theta_2 =0}.
 \end{eqnarray}
where $b_1$ and $b_2$ are Nakanishi-Lautrup type auxiliary fields.
The ghost term corresponding to this gauge fixing term can be written as 
  \begin{eqnarray}
   \mathcal{L}^{1+ 2+}_{gh} &=&    (D^2_1 + \partial_3) [ \bar{c}_1  D^a_1\nabla_{1a}    c_1 ]_{\theta_1 =0} \nonumber \\
   && + (D^2_2 + \partial_3) 
   [  \bar{c}_2  D^a_2\nabla_{2a}    c_2 ]_{\theta_2 =0}
 ,
\nonumber \end{eqnarray}
\begin{eqnarray}
   \mathcal{L}^{1- 2-}_{gh} &=&    (D^2_1 - \partial_3) [ \bar{c}_1  D^a_1\nabla_{1a}    c_1]_{\theta_1 =0} 
   \nonumber \\ &&  + (D^2_2 - \partial_3) [  \bar{c}_2  D^a_2\nabla_{2a}    c_2]_{\theta_2 =0}
 ,
\nonumber \end{eqnarray}
\begin{eqnarray}
   \mathcal{L}^{1+ 2-}_{gh} &=&  
   (D^2_1 + \partial_3) [ \bar{c}_1  D^a_1\nabla_{1a}    c_1]_{\theta_1 =0}  \nonumber \\ && + (D^2_2 - \partial_3) [ \bar{c}_2  D^a_2\nabla_{2a}    c_2]_{\theta_2 =0}
 ,
\nonumber \end{eqnarray}
\begin{eqnarray}
   \mathcal{L}^{1- 2+}_{gh} &=&  
  (D^2_1- \partial_3) [ \bar{c}_1  D^a_1\nabla_{1a}    c_1]_{\theta_1 =0}  \nonumber \\ &&  
  + (D^2_2 + \partial_3) [  \bar{c}_2  D^a_2\nabla_{2a}    c_2]_{\theta_2 =0}
 \end{eqnarray}
 where $c_1, {c}_2$ are the    ghost fields and $\bar c_1,  \bar{c}$  are the anti-ghost fields. 
Now we can define $\mathcal{L}^{1\pm 2\pm}_g$ as 
  \begin{eqnarray}
   \mathcal{L}^{1+ 2+}_{g} &=&     \mathcal{L}^{1\pm 2\pm}_{gf} +    \mathcal{L}^{1\pm  2\pm }_{gh}. 
 \end{eqnarray}
 The effective Lagrangian    $ {\cal L}^{1\pm 2\pm}={\cal L}^{1\pm 2\pm} + {\cal L}_{g}^{1\pm 2\pm}$ 
 which is given by the sum of the ghost and the gauge fixing terms with 
 modified super-Yang-Mills Lagrangian is invariant under the following BRST transformations
\begin{eqnarray}
s_b \,\Gamma_{1a} = \nabla_{1a}    c_1, && s_b \, \Gamma_{2a} =\nabla_{2a}    
 c_2, \nonumber \\
s_b  \,c_{1} = -  \frac{1}{2} {[c_{1},c_{1}]}_ { }, && s_b  \,{ {c_{2}}} = 
-  \frac{1}{2} [{ {c_{2}}} ,   c_{2}]_{ }, \nonumber \\ 
s_b  \,\bar{c}_{1} =  b_1, && s_b  \,{\bar c}_{2} =   b_{2}, \nonumber \\
s_b  \, b_1 =0, && s_b  \,  b_{2}= 0,  
\end{eqnarray}
This is because modified super-Yang-Mills Lagrangian is BRST invariant, and the sum of the 
  gauge fixing and ghost terms can be expressed as  
    \begin{eqnarray}
   \mathcal{L}^{1+ 2+}_{g} &=&   s_b (D^2_1 + \partial_3) [  \bar c_1  D^a_1 \Gamma_{1a} +  \frac{\alpha}{2} \bar c_1 {b}_1 ]_{\theta_1 =0} 
   \nonumber \\ && + s_b (D^2_2 + \partial_3) 
   [   \bar c_2  D^a_2 \Gamma_{2a} +  \frac{\alpha}{2} \bar c_2 {b}_2 ]_{\theta_2 =0}
 ,
\nonumber \end{eqnarray}
\begin{eqnarray}
   \mathcal{L}^{1- 2-}_{g} &=&   s_b (D^2_1 - \partial_3) [  \bar c_1  D^a_1 \Gamma_{1a} +  \frac{\alpha}{2} \bar c_1 {b}_1]_{\theta_1 =0} 
   \nonumber \\ &&  + s_b (D^2_2 - \partial_3) [   \bar c_2  D^a_2 \Gamma_{2a} +  \frac{\alpha}{2} \bar c_2 {b}_2]_{\theta_2 =0}
 ,
\nonumber \end{eqnarray}
\begin{eqnarray}
   \mathcal{L}^{1+ 2-}_{g} &=&  
   s_b (D^2_1 + \partial_3) [  \bar c_1  D^a_1 \Gamma_{1a} +  \frac{\alpha}{2} \bar c_1 {b}_1]_{\theta_1 =0}  \nonumber \\
   && + s_b (D^2_2 - \partial_3) [  \bar c_2  D^a_2 \Gamma_{2a} +  \frac{\alpha}{2} \bar c_2 {b}_2]_{\theta_2 =0}
 ,
\nonumber \end{eqnarray}
\begin{eqnarray}
   \mathcal{L}^{1- 2+}_{g} &=&  
 s_b  (D^2_1- \partial_3) [  \bar c_1  D^a_1 \Gamma_{1a} +  \frac{\alpha}{2} \bar c_1 {b}_1]_{\theta_1 =0}  \nonumber \\ &&  
  + s_b (D^2_2 + \partial_3) [   \bar c_2  D^a_2 \Gamma_{2a} +  \frac{\alpha}{2} \bar c_2 {b}_2]_{\theta_2 =0}.
 \end{eqnarray}
 Now as $s_b^2 = 0$, the sum of the gauge fixing and ghost terms  is also invariant under the BRST transformations. 

It is possible to analyze this theory with fixed background fields, and quantum fluctuations around these fields. We 
can obtain the BRST symmetry of such a theory. The Lagrangian is expressed in terms of classical background fields  
and quantum fluctuations around these fields,
\begin{eqnarray}
 \mathcal{L}^{1\pm 2+\pm} (\Gamma_1, \Gamma_2) + \mathcal{L}_g^{1\pm 2+\pm} (\Gamma_1, \Gamma_2, c_1, c_2, \bar c_1, \bar c_2, b_1, b_2)\to  \nonumber 
 \\
 \mathcal{L}_g^{1\pm 2+\pm} (\Gamma_1-\tilde \Gamma_1, \Gamma_2-\tilde \Gamma_2, c_1-\tilde c_1, c_2-\tilde c_2, \bar c_1-\tilde{\bar c}_1, c_2-\tilde{\bar c}_2, 
 b_1-\tilde b_1, b_2-\tilde b_2)  
 \nonumber \\ + 
 \mathcal{L}^{1\pm 2+\pm} (\Gamma_1- \tilde \Gamma_1 , \Gamma_2-\tilde \Gamma_2). 
\end{eqnarray}
Here the fields $\tilde \Gamma_{1a}, \tilde \Gamma_{2a},  \tilde c_1, \tilde c_2, \tilde{\bar c}_1, \tilde{\bar c}_2, 
 \tilde b_1, \tilde b_2$ are quantum fluctuations around the background fields. 
Let us express the fields as a sum of the background fields and quantum fluctuations around them, 
\begin{eqnarray}
 \Gamma_{1a} \to \Gamma_{1a} + \tilde{\Gamma}_{1a}, && 
 \Gamma_{2a} \to \Gamma_{2a} + \tilde{\Gamma}_{2a},\nonumber \\
 c_1 \to c_1 + \tilde{c}_1, && 
  c_2 \to c_2 + \tilde{c}_2,
  \nonumber \\
  \bar{c}_1 \to \bar{c}_1 + \tilde{\bar{c}}_1, &&
    \bar{c}_2 \to \bar{c}_2 + \tilde{\bar{c}}_2,
    \nonumber \\
    b_1 \to b_1 + \tilde{b}_1, && 
    b_2 \to b_2 + \tilde{b}_2. 
\end{eqnarray}
So, the covariant derivative transforms to  $\nabla_{1a} \to \bar\nabla_{1a} = D_{a} - i \Gamma_{1a} -i \tilde\Gamma_{1a}$ and 
$\nabla_{2a} \to \bar\nabla_{2a} = D_{a} - i \Gamma_{2a} -i \tilde\Gamma_{2a}$ 
 Now the quantum fluctuations transform as follows, 
 \begin{eqnarray}
s_b \,\tilde\Gamma_{1a} = \psi_{1a} - \bar\nabla_{1a}    (c_1- \tilde c_1), 
&& s_b \,\tilde \Gamma_{2a} =\psi_{2a} -\bar \nabla_{2a}    
 (c_2-\tilde c_2), \nonumber\\ 
s_b  \,\tilde c_{1} = \lambda_1 -  \frac{1}{2} {[c_{1}- \tilde c_1,c_{1} -\tilde c_1]}_ { },
&& s_b  \,{ \tilde{c_{2}}} = \lambda_2 -  \frac{1}{2} [{ {c_{2}}} -\tilde c_2,  
c_{2}-\tilde c_2]_{ }, \nonumber  \\
s_b  \,\tilde{\bar{c}}_{1} = \bar \lambda_1 - (b_1 -\tilde b_1), && s_b  \,\tilde{\bar c}_{2} = 
\bar \lambda_2 - (b_{2}- \tilde b_2), \nonumber \\ 
s_b  \, \tilde b_1 =\mu_1, && s_b  \,  \tilde b_{2}=\mu_2.
\end{eqnarray} 
The BRST symmetry for the background fields can be expressed as 
\begin{eqnarray}
s_b \,\Gamma_{1a} = \psi_{1a}, && s_b \, \Gamma_{2a} =\psi_{2a}, \nonumber \\
s_b  \,c_{1} = \lambda_{1a}, && s_b  \,{ {c_{2}}} = \lambda_{2a}, \nonumber \\ 
s_b  \,\bar{c}_{1} =  \bar \lambda_1, && s_b  \,{\bar c}_{2} =   \bar \lambda_{2}, \nonumber \\
s_b  \, b_1 =\mu_1, && s_b  \,  b_{2}= \mu_2.   
\end{eqnarray}
Here we have introduced new  ghosts associated with the shift symmetry,  and the BRST 
transformation of these new ghost fields vanishes 
$s_b \psi_{1a} = s_b \psi_{2a} =0$, and $s_b \lambda_1 = s_b \lambda_2 = s_b \bar \lambda_1 =
s_b \bar \lambda_2 = s_b \mu_1 = s_b \bar \mu_2 =0$. Now we double the field content of this theory 
by adding a set of anti-fields corresponding to each field, and the BRST transformation of 
these anti fields is  given by 
\begin{eqnarray}
s_b \,\Gamma_{1a}^* = u_{1a}, && s_b \, \Gamma_{2a}^* =u_{2a}, \nonumber \\
s_b  \,c_{1}^* = v_{1}, && s_b  \,{ {c_{2}}}^* = v_{2}, \nonumber \\ 
s_b  \,\bar{c}_{1}^* =  \bar v_1, && s_b  \,{\bar c}_{2}^* =   \bar v_{2}, \nonumber \\
s_b  \, b_1^* =t_1, && s_b  \,  b_{2}^*= t_2.    
\end{eqnarray}
Finally, the BRST transformation of these auxiliary fields vanish, 
$ s_b u_{1a} = s_b u_{2a} =0$ and $ s_b v_1 = s_b v_2 = s_b \bar v_1 = s_b \bar v_2 =
s_b t_1 = s_b t_2 =0$.  

Now we can add the following term to the sum of the gauge fixing term and ghost term, 
    \begin{eqnarray}
   \mathcal{L}^{1+ 2+}_{f} &=&    (D^2_1 + \partial_3) [   \Gamma^{1a *} s_b \Gamma_{1a} - c^*_1 s_b  c_1]_{\theta_1 =0} 
   \nonumber \\ && +  (D^2_2 + \partial_3) 
   [    \Gamma^{2a *} s_b \Gamma_{2a} - c^*_2 s_b  c_2 ]_{\theta_2 =0}
 ,
\nonumber \end{eqnarray}
\begin{eqnarray}
   \mathcal{L}^{1- 2-}_{f} &=&    (D^2_1 - \partial_3) [   \Gamma^{1a *} s_b \Gamma_{1a} - c^*_1 s_b  c_1]_{\theta_1 =0} 
   \nonumber \\ &&  +  (D^2_2 - \partial_3) [    \Gamma^{2a *} s_b \Gamma_{2a} - c^*_2 s_b  c_2]_{\theta_2 =0}
 ,
\nonumber \end{eqnarray}
\begin{eqnarray}
   \mathcal{L}^{1+ 2-}_{f} &=&  
    (D^2_1 + \partial_3) [   \Gamma^{1a *} s_b \Gamma_{1a} - c^*_1 s_b  c_1]_{\theta_1 =0}  \nonumber \\
   && +  (D^2_2 - \partial_3) [   \Gamma^{2a *} s_b \Gamma_{2a} - c^*_2 s_b  c_2]_{\theta_2 =0}
 ,
\nonumber \end{eqnarray}
\begin{eqnarray}
   \mathcal{L}^{1- 2+}_{f} &=&  
   (D^2_1- \partial_3) [   \Gamma^{1a *} s_b \Gamma_{1a} - c^*_1 s_b  c_1]_{\theta_1 =0}  \nonumber \\ &&  
  +  (D^2_2 + \partial_3) [    \Gamma^{2a *} s_b \Gamma_{2a} - c^*_2 s_b  c_2]_{\theta_2 =0}.
 \end{eqnarray}
Now we can write the total action for this theory as 
\begin{eqnarray}
 \Gamma^{1+ 2+ 0} = \int d^3 x [ \mathcal{L}^{1\pm 2\pm} + \mathcal{L}^{1\pm 2\pm}_g + 
 \mathcal{L}^{1\pm 2\pm}_f].  
\end{eqnarray}
Then we can calculate the effective action, and to the first order term that corresponds to
this classical action.
We  can write the  Slavnov-Taylor Identity for this theory as  
    \begin{eqnarray}
  && \int d^3 x   (D^2_1 + \partial_3) \nonumber \\ && \times \left[  \frac{\delta  \Gamma^{1+ 2+ 0}}{\delta \Gamma_{1a}^* }\frac{\delta  \Gamma^{1+ 2+ 0}}{\delta \Gamma_{1a} } + 
\frac{\delta  \Gamma^{1+ 2+ 0}}{\delta c^*_1 }\frac{\delta  \Gamma^{1+ 2+ 0}}{\delta c_1 } + 
b_1 \frac{\delta  \Gamma^{1+ 2+ 0}}{\delta \bar c_1 }\right]_{\theta_1 =0} 
   \nonumber \\ && +  \int d^3 x  (D^2_2 + \partial_3) \nonumber \\ && \times 
   \left[   \frac{\delta  \Gamma^{1+ 2+ 0}}{\delta \Gamma_{2a}^* }\frac{\delta  \Gamma^{1+ 2+ 0}}{\delta \Gamma_{2a} } + 
\frac{\delta  \Gamma^{1+ 2+ 0}}{\delta c^*_2 }\frac{\delta  \Gamma^{1+ 2+ 0}}{\delta c_2 } + 
b_2 \frac{\delta  \Gamma^{1+ 2+ 0}}{\delta \bar c_2 } \right]_{\theta_2 =0} \nonumber \\ &=&0 
 ,
\nonumber \end{eqnarray}
\begin{eqnarray}
   && \int d^3 x (D^2_1 - \partial_3)\nonumber \\ && \times  \left[   \frac{\delta  \Gamma^{1+ 2+ 0}}{\delta \Gamma_{1a}^* }\frac{\delta  \Gamma^{1+ 2+ 0}}{\delta \Gamma_{1a} } + 
\frac{\delta  \Gamma^{1+ 2+ 0}}{\delta c^*_1 }\frac{\delta  \Gamma^{1+ 2+ 0}}{\delta c_1 } + 
b_1 \frac{\delta  \Gamma^{1+ 2+ 0}}{\delta \bar c_1 }\right]_{\theta_1 =0} 
   \nonumber \\ &&  + \int d^3 x  (D^2_2 - \partial_3) \nonumber \\ && \times \left[   \frac{\delta  \Gamma^{1+ 2+ 0}}{\delta \Gamma_{2a}^* }\frac{\delta  \Gamma^{1+ 2+ 0}}{\delta \Gamma_{2a} } + 
\frac{\delta  \Gamma^{1+ 2+ 0}}{\delta c^*_2 }\frac{\delta  \Gamma^{1+ 2+ 0}}{\delta c_2 } + 
b_2 \frac{\delta  \Gamma^{1+ 2+ 0}}{\delta \bar c_2 }\right]_{\theta_2 =0} \nonumber \\ &=&0
 ,
\nonumber \end{eqnarray}
\begin{eqnarray}
  &&
    \int d^3 x  (D^2_1 - \partial_3) \nonumber \\ && \times \left[   \frac{\delta  \Gamma^{1+ 2+ 0}}{\delta \Gamma_{1a}^* }\frac{\delta  \Gamma^{1+ 2+ 0}}{\delta \Gamma_{1a} } + 
\frac{\delta  \Gamma^{1+ 2+ 0}}{\delta c^*_1 }\frac{\delta  \Gamma^{1+ 2+ 0}}{\delta c_1 } + 
b_1 \frac{\delta  \Gamma^{1+ 2+ 0}}{\delta \bar c_1 }\right]_{\theta_1 =0}  \nonumber \\
   && +  \int d^3 x (D^2_2 + \partial_3)\nonumber \\ && \times  \left[  \frac{\delta  \Gamma^{1+ 2+ 0}}{\delta \Gamma_{2a}^* }\frac{\delta  \Gamma^{1+ 2+ 0}}{\delta \Gamma_{2a} } + 
\frac{\delta  \Gamma^{1+ 2+ 0}}{\delta c^*_2 }\frac{\delta  \Gamma^{1+ 2+ 0}}{\delta c_2 } + 
b_2 \frac{\delta  \Gamma^{1+ 2+ 0}}{\delta \bar c_2 }\right]_{\theta_2 =0} \nonumber \\ &=&0
 ,
\nonumber \end{eqnarray}
\begin{eqnarray}  
  && \int d^3 x  (D^2_1 + \partial_3) \nonumber \\ && \times \left[ \frac{\delta  \Gamma^{1+ 2+ 0}}{\delta \Gamma_{1a}^* }\frac{\delta  \Gamma^{1+ 2+ 0}}{\delta \Gamma_{1a} } + 
\frac{\delta  \Gamma^{1+ 2+ 0}}{\delta c^*_1 }\frac{\delta  \Gamma^{1+ 2+ 0}}{\delta c_1 } + 
b_1 \frac{\delta  \Gamma^{1+ 2+ 0}}{\delta \bar c_1 }\right]_{\theta_1 =0}  \nonumber \\ &&  
  +  \int d^3 x  (D^2_2 - \partial_3) \nonumber \\ && \times \left[   \frac{\delta  \Gamma^{1+ 2+ 0}}{\delta \Gamma_{2a}^* }\frac{\delta  \Gamma^{1+ 2+ 0}}{\delta \Gamma_{2a} } + 
\frac{\delta  \Gamma^{1+ 2+ 0}}{\delta c^*_2 }\frac{\delta  \Gamma^{1+ 2+ 0}}{\delta c_2 } + 
b_2 \frac{\delta  \Gamma^{1+ 2+ 0}}{\delta \bar c_2 }\right]_{\theta_2 =0} \nonumber \\ &=&0.
 \end{eqnarray}
This procedure can be followed by using the effective action to obtain Slavnov-Taylor Identity at higher order. 
It may be noted that the tree level Slavnov-Taylor Identity can be used to relate relating the two, three and four point functions. 
This has been used for analyzing the consistency of occurring at one loop in noncommutative gauge theories \cite{nonc}. 
It will be possible to use a similar analysis here and analyze the divergences occurring in the supersymmetric Yang-Mills theory. 
However, the most important observation of this analysis is that the standard form of the 
Slavnov-Taylor Identity  does not get deformed, and it is only the measure that is deformed for such theories. 
These Slavnov-Taylor Identity depend on the gauge symmetry of the theory, and the gauge symmetry of the theory 
is not broken in Yang-Mills by the presence of a boundary.

 \section{Boundary Action}
 In this section, we will analyse the boundary action by using  the projection operators,   
  $P_{\pm} = (1 \pm \gamma^3)/2$.
We can project the superderivatives using these projection operators
 as,
 $D_{1 \pm a} = (P_\pm)_{a}^{\;b} D_{1b}$ and $D_{2 \pm a} = (P_\pm)_{a}^{\;b} D_{2b}$.
 The supercharges can also be projected as  $Q_{1 \pm a} = (P_\pm)_{a}^{\;b} Q_{1b}$ and 
 $Q_{2 \pm a} = (P_\pm)_{a}^{\;b} Q_{2b}$ \cite{1}. 
The     bulk supercharges $Q_{1a}$ and $Q_{2a}$ can now be expressed as  \cite{4}
\begin{eqnarray}
  \epsilon^{1a} Q_{1a} &=&   \epsilon^{1a}( P_- + P_+) Q_{1a}  \nonumber \\
  &=&
  \epsilon^{1+} Q_{1-} + \epsilon^{1-} Q_{1+},\nonumber \end{eqnarray}
\begin{eqnarray}
 \epsilon^{2a} Q_{2a} &=& \epsilon^{2a}( P_- + P_+) Q_{2a} \nonumber \\
 &=&
  \epsilon^{2+} Q_{2-} + \epsilon^{2-} Q_{2+}.
\end{eqnarray}
These bulk supercharges $Q_{1\pm},    Q_{2\pm},  $ are related to the 
boundary supercharges $Q'_{1\pm},   Q'_{2\pm },  $ as 
\begin{eqnarray}
 Q_{1-} = Q'_{1- }+ \theta_{1-} \partial_3, &&
 Q_{1+} = Q'_{1+} - \theta_{1+}\partial_3,  \nonumber \end{eqnarray}
\begin{eqnarray}
  Q_{2-}= Q'_{2- }+ \theta_{2-} \partial_3, &&
 Q_{2+} = Q'_{2+} - \theta_{2+}\partial_3,
\end{eqnarray}
Here the  boundary
supercharges are defined as  
\begin{eqnarray}
  Q'_{1+} = \partial_{1+} - \gamma^s \theta_{1-} \partial_s, &&
   Q'_{1-} = \partial_{1-} - \gamma^s \theta_{1+} \partial_s,\nonumber \end{eqnarray}
\begin{eqnarray}
  Q'_{2+} = \partial_{2+} - \gamma^s \theta_{2-} \partial_s, &&
   Q'_{2-} = \partial_{2-} - \gamma^s \theta_{2+} \partial_s,
\end{eqnarray}
where $s$
is the
index for the coordinates along the boundary,  i.e., 
the case $\mu = 3$ has been excluded for a boundary fixed at $x_3$.
The supercharges  $Q_{1\pm }$ and $  Q_{2\pm}  $ are the generators of the
half supersymmetry for the bulk fields. Furthermore, 
$Q'_{1\pm} $ and $  Q'_{2\pm}   $ are the standard
generators of the supersymmetry for  the boundary fields.
It is possible to express the boundary supercharges as \cite{8} 
\begin{eqnarray}
 Q'_{1-} &=& \exp ( + \theta_{1+} \theta_{1-} \partial_3)
 Q_{1-} \exp ( - \theta_{1+} \theta_{1-} \partial_3),
\nonumber \end{eqnarray}
\begin{eqnarray}
 Q'_{1+} &=& \exp ( -  \theta_{1-} \theta_{1+} \partial_3)
 Q_{1+}\exp ( +  \theta_{1-} \theta_{1+} \partial_3),
 \nonumber \end{eqnarray}
\begin{eqnarray}
  Q'_{2-} &=& \exp ( + \theta_{2+} \theta_{2-} \partial_3)
 Q_{2-}\exp ( - \theta_{2+} \theta_{2-} \partial_3),
\nonumber \end{eqnarray}
\begin{eqnarray}
 Q'_{2+} &=& \exp ( -  \theta_{2-} \theta_{2+} \partial_3)
 Q_{2+}\exp ( +  \theta_{2-} \theta_{2+} \partial_3).
\end{eqnarray}
It is also possible to write the super-algebra of the bulk supercharges in presence of a boundary as 
\begin{eqnarray}
 \{Q_{1+ a}, Q_{1+ b}\} = 2 (\gamma_{ab}^{s}P_+)\partial_s   ,
 &&  \{D_{1+a}, D_{1+b}\} =- 2 (\gamma_{ab}^{s}P_+)\partial_s  , \nonumber \end{eqnarray}
\begin{eqnarray}
 \{Q_{1- a}, Q_{1- b}\} = 2 (\gamma_{ab}^{s}P_-)\partial_s   ,
 &&  \{D_{1-a}, D_{1-b}\} =- 2 (\gamma_{ab}^{s}P_-)\partial_s   , \nonumber \end{eqnarray}
\begin{eqnarray}
 \{Q_{1+a}, Q_{1-b}\} = -2 (P_{-})_{ab}\partial_3   ,
 &&  \{D_{1+a}, D_{1-b}\} = 2 (P_-)_{ab}\partial_3   , \nonumber \end{eqnarray}
\begin{eqnarray}
 \{Q_{2  + a}, Q_{2  + b}\} = 2 (\gamma_{ab}^{s}P_+)\partial_s   ,
 &&  \{D_{2  +a}, D_{2  +b}\} =- 2 (\gamma_{ab}^{s}P_+)\partial_s  , \nonumber \end{eqnarray}
\begin{eqnarray}
 \{Q_{2  - a}, Q_{2  - b}\} = 2 (\gamma_{ab}^{s}P_-)\partial_s   ,
 &&  \{D_{2  -a}, D_{2  -b}\} =- 2 (\gamma_{ab}^{s}P_-)\partial_s   , \nonumber \end{eqnarray}
\begin{eqnarray}
 \{Q_{2  +a}, Q_{2  -b}\} = -2 (P_{-})_{ab}\partial_3   ,
 &&  \{D_{2  +a}, D_{2  -b}\} = 2 (P_-)_{ab}\partial_3 .
\end{eqnarray}
It may be noted that $\{Q_{1\pm}, Q_{2\pm}\} = \{D_{1\pm}, D_{2\pm}\} =0$, and
$\{Q_{1\pm}, D_{2\pm}\} = \{Q_{1\pm}, D_{1\pm}\} =\{Q_{2\pm}, D_{2\pm}\} = \{Q_{2\pm}, D_{1\pm}\} =0$.
Thus, we can write 
\begin{eqnarray}
 D_{1-a}D_{1+b} = (P_-)_{ab} (\partial_3 -D_1^2), &&
 D_{1+a}D_{1-b} =  -(P_-)_{ab} (\partial_3 + D_1^2),
  \nonumber \end{eqnarray}
\begin{eqnarray}
   D_{2-a}D_{2+b} = (P_-)_{ab} (\partial_3 -D_2^2), &&
 D_{2+a}D_{2-b} = -(P_-)_{ab} (\partial_3 + D_2^2).
\end{eqnarray}
Contracting these equation and using $(P_-)_a^a =1$, we  obtained \cite{8},
\begin{eqnarray}
D_1^2 + \partial_3 =  D_{1+}D_{1-}, && D_2^2 + \partial_3 =D_{2+}D_{2-}, \label{a1} \\
D_1^2 - \partial_3 = D_{1-}D_{1+},&& D_2^2 - \partial_3 =D_{2-}D_{2+}.\label{a4}
\end{eqnarray}

We can write the Lagrangian for the super-Yang-Mills theory in presence of a boundary as 
 \begin{eqnarray}
   \mathcal{L}^{1+ 2+} &=&    D_{1+}D_{1-} [W^a_1W_{1a}]_{\theta_1 =0} + D_{2+}D_{2-} [ W^a_2W_{2a}]_{\theta_2 =0}
 ,
\nonumber \end{eqnarray}
\begin{eqnarray}
   \mathcal{L}^{1- 2-} &=&    D_{1-}D_{1+} [W^a_1W_{1a}]_{\theta_1 =0} + D_{2-}D_{2+} [ W^a_2W_{2a}]_{\theta_2 =0}
 ,
\nonumber \end{eqnarray}
\begin{eqnarray}
   \mathcal{L}^{1+ 2-} &=&  
   D_{1-}D_{1+} [W^a_1W_{1a}]_{\theta_1 =0} + D_{2+}D_{2-} [ W^a_2W_{2a}]_{\theta_2 =0}
 ,
\nonumber \end{eqnarray}
\begin{eqnarray}
   \mathcal{L}^{1- 2+} &=&  
  D_{1+}D_{1-} [W^a_1W_{1a}]_{\theta_1 =0} + D_{2-}D_{2+} [ W^a_2W_{2a}]_{\theta_2 =0}.
 \end{eqnarray}
 We can now write the gauge fixing terms in the Lorentz gauge as 
  \begin{eqnarray}
   \mathcal{L}^{1+ 2+}_{gf} &=&    D_{1+}D_{1-} [ b_1   (D^a_1 \Gamma_{1a}) + \frac{\alpha}{2}b^2_1 ]_{\theta_1 =0} \nonumber  
\\ && + D_{2+}D_{2-} 
   [  b_2  (D^a_2 \Gamma_{2a}) + \frac{\alpha}{2}b^2 ]_{\theta_2 =0}
 ,
\nonumber \end{eqnarray}
\begin{eqnarray}
   \mathcal{L}^{1- 2-}_{gf} &=&    D_{1-}D_{1+} [ b_1   (D^a_1 \Gamma_{1a}) + \frac{\alpha}{2}b^2_1]_{\theta_1 =0} 
   \nonumber \\  &&  + D_{2-}D_{2+} [  b_2  (D^a_2 \Gamma_{2a}) + \frac{\alpha}{2}b^2]_{\theta_2 =0}
 ,
\nonumber \end{eqnarray}
\begin{eqnarray}
   \mathcal{L}^{1+ 2-}_{gf} &=&  
   D_{1-}D_{1+} [ b_1   (D^a_1 \Gamma_{1a}) + \frac{\alpha}{2}b^2_1]_{\theta_1 =0}  \nonumber \\
   && + D_{2+}D_{2-} [ b_2  (D^a_2 \Gamma_{2a}) + \frac{\alpha}{2}b^2]_{\theta_2 =0}
 ,
\nonumber \end{eqnarray}
\begin{eqnarray}
   \mathcal{L}^{1- 2+}_{gf} &=&  
  D_{1+}D_{1-} [ b_1   (D^a_1 \Gamma_{1a}) + \frac{\alpha}{2}b^2_1]_{\theta_1 =0}  \nonumber \\ &&  
  + D_{2-}D_{2+} [  b_2  (D^a_2 \Gamma_{2a}) + \frac{\alpha}{2}b^2]_{\theta_2 =0}.
 \end{eqnarray}
 The ghost terms corresponding to this gauge fixing term can   be written as 
   \begin{eqnarray}
   \mathcal{L}^{1+ 2+}_{gh} &=&    D_{1+}D_{1-} [ \bar{c}_1  D^a_1\nabla_{1a}    c_1 ]_{\theta_1 =0} \nonumber \\
  && + D_{2+}D_{2-} 
   [  \bar{c}_2  D^a_2\nabla_{2a}    c_2 ]_{\theta_2 =0}
 ,
\nonumber \end{eqnarray}
\begin{eqnarray}
   \mathcal{L}^{1- 2-}_{gh} &=&    D_{1-}D_{1+} [ \bar{c}_1  D^a_1\nabla_{1a}    c_1]_{\theta_1 =0} 
   \nonumber \\ &&  + D_{2-}D_{2+} [  \bar{c}_2  D^a_2\nabla_{2a}    c_2]_{\theta_2 =0}
 ,
\nonumber \end{eqnarray}
\begin{eqnarray}
   \mathcal{L}^{1+ 2-}_{gh} &=&  
   D_{1-}D_{1+} [ \bar{c}_1  D^a_1\nabla_{1a}    c_1]_{\theta_1 =0}  \nonumber \\ 
   && + D_{2+}D_{2-} [ \bar{c}_2  D^a_2\nabla_{2a}    c_2]_{\theta_2 =0}
 ,
\nonumber \end{eqnarray}
\begin{eqnarray}
   \mathcal{L}^{1- 2+}_{gh} &=&  
  D_{1+}D_{1-} [ \bar{c}_1  D^a_1\nabla_{1a}    c_1]_{\theta_1 =0}  \nonumber \\  &&  
  + D_{2-}D_{2+} [  \bar{c}_2  D^a_2\nabla_{2a}    c_2]_{\theta_2 =0}.
 \end{eqnarray}
The total effective Lagrangian which is given by a sum of the gauge fixing term and  the ghost term with 
the original Lagrangian can be written as 
\begin{eqnarray}
     \mathcal{L}^{1+ 2+} +  \mathcal{L}^{1+ 2+}_{g} &=&   s_b D_{1+}D_{1-} [  \bar c_1  D^a_1 \Gamma_{1a} +  \frac{\alpha}{2} \bar c_1 {b}_1 ]_{\theta_1 =0}
   \nonumber \\ && + s_b D_{2+}D_{2-} 
   [   \bar c_2  D^a_2 \Gamma_{2a} +  \frac{\alpha}{2} \bar c_2 {b}_2 ]_{\theta_2 =0}
 \nonumber \\  &&  +
  D_{1+}D_{1-} [W^a_1W_{1a}]_{\theta_1 =0} + D_{2+}D_{2-} [ W^a_2W_{2a}]_{\theta_2 =0},
\nonumber \end{eqnarray}
\begin{eqnarray}
    \mathcal{L}^{1- 2-} +   \mathcal{L}^{1- 2-}_{g} &=&   s_b D_{1-}D_{1+} [  \bar c_1  D^a_1 \Gamma_{1a} +  \frac{\alpha}{2} \bar c_1 {b}_1]_{\theta_1 =0} 
   \nonumber \\ &&  + s_b D_{2-}D_{2+} [   \bar c_2  D^a_2 \Gamma_{2a} +  \frac{\alpha}{2} \bar c_2 {b}_2]_{\theta_2 =0}
 \nonumber  \\&& +
  D_{1-}D_{1+} [W^a_1W_{1a}]_{\theta_1 =0} + D_{2-}D_{2+} [ W^a_2W_{2a}]_{\theta_2 =0},
\nonumber \end{eqnarray}
\begin{eqnarray}
    \mathcal{L}^{1+ 2-} +   \mathcal{L}^{1+ 2-}_{g} &=&  
   s_b D_{1-}D_{1+} [  \bar c_1  D^a_1 \Gamma_{1a} +  \frac{\alpha}{2} \bar c_1 {b}_1]_{\theta_1 =0}  \nonumber \\ 
   && + s_b D_{2+}D_{2-} [  \bar c_2  D^a_2 \Gamma_{2a} +  \frac{\alpha}{2} \bar c_2 {b}_2]_{\theta_2 =0}
 \nonumber  \\ && +
    D_{1-}D_{1+} [W^a_1W_{1a}]_{\theta_1 =0} + D_{2+}D_{2-} [ W^a_2W_{2a}]_{\theta_2 =0},
\nonumber \end{eqnarray}
\begin{eqnarray}
  \mathcal{L}^{1- 2+} +   \mathcal{L}^{1- 2+}_{g} &=&  
 s_b  D_{1+}D_{1-} [  \bar c_1  D^a_1 \Gamma_{1a} +  \frac{\alpha}{2} \bar c_1 {b}_1]_{\theta_1 =0}  \nonumber \\ &&   
  + s_b D_{2-}D_{2+} [   \bar c_2  D^a_2 \Gamma_{2a} +  \frac{\alpha}{2} \bar c_2 {b}_2]_{\theta_2 =0} \nonumber  \\ && + 
  D_{1+}D_{1-} [W^a_1W_{1a}]_{\theta_1 =0} + D_{2-}D_{2+} [ W^a_2W_{2a}]_{\theta_2 =0}.\nonumber \\ 
 \end{eqnarray}
  Now we can  write $\mathcal{L}^{1\pm 2\pm}_f$ as, 
    \begin{eqnarray}
   \mathcal{L}^{1+ 2+}_{f} &=&    D_{1+} D_{1-} [   \Gamma^{1a *} s_b \Gamma_{1a} - c^*_1 s_b  c_1]_{\theta_1 =0} 
   \nonumber \\ && +  D_{2+} D_{2-}
   [    \Gamma^{2a *} s_b \Gamma_{2a} - c^*_2 s_b  c_2 ]_{\theta_2 =0}
 ,
\nonumber \end{eqnarray}
\begin{eqnarray}
   \mathcal{L}^{1- 2-}_{f} &=&   D_{1-} D_{1+} [   \Gamma^{1a *} s_b \Gamma_{1a} - c^*_1 s_b  c_1]_{\theta_1 =0} 
   \nonumber \\ &&  + D_{2-} D_{2+} [    \Gamma^{2a *} s_b \Gamma_{2a} - c^*_2 s_b  c_2]_{\theta_2 =0}
 ,
\nonumber \end{eqnarray}
\begin{eqnarray}
   \mathcal{L}^{1+ 2-}_{f} &=&  
    D_{1+} D_{1-} [   \Gamma^{1a *} s_b \Gamma_{1a} - c^*_1 s_b  c_1]_{\theta_1 =0}  \nonumber \\
   && +  D_{2-} D_{2+} [   \Gamma^{2a *} s_b \Gamma_{2a} - c^*_2 s_b  c_2]_{\theta_2 =0}
 ,
\nonumber \end{eqnarray}
\begin{eqnarray}
   \mathcal{L}^{1- 2+}_{f} &=&  
   D_{1-} D_{1} [   \Gamma^{1a *} s_b \Gamma_{1a} - c^*_1 s_b  c_1]_{\theta_1 =0}  \nonumber \\ &&  
  +  D_{2+} D_{2-} [    \Gamma^{2a *} s_b \Gamma_{2a} - c^*_2 s_b  c_2]_{\theta_2 =0}.
 \end{eqnarray}
So, the   Slavnov-Taylor Identity for this theory as  
    \begin{eqnarray}
  &&  \int d^3 x  D_{1+} D_{1-} \nonumber \\ && \times  \left[  \frac{\delta  \Gamma^{1+ 2+ 0}}{\delta \Gamma_{1a}^* }\frac{\delta  \Gamma^{1+ 2+ 0}}{\delta \Gamma_{1a} } + 
\frac{\delta  \Gamma^{1+ 2+ 0}}{\delta c^*_1 }\frac{\delta  \Gamma^{1+ 2+ 0}}{\delta c_1 } + 
b_1 \frac{\delta  \Gamma^{1+ 2+ 0}}{\delta \bar c_1 }\right]_{\theta_1 =0} 
   \nonumber \\ && + \int d^3 x  D_{2+} D_{2-} \nonumber \\ && \times 
   \left[   \frac{\delta  \Gamma^{1+ 2+ 0}}{\delta \Gamma_{2a}^* }\frac{\delta  \Gamma^{1+ 2+ 0}}{\delta \Gamma_{2a} } + 
\frac{\delta  \Gamma^{1+ 2+ 0}}{\delta c^*_2 }\frac{\delta  \Gamma^{1+ 2+ 0}}{\delta c_2 } + 
b_2 \frac{\delta  \Gamma^{1+ 2+ 0}}{\delta \bar c_2 } \right]_{\theta_2 =0} \nonumber \\ &=&0 
 ,
\nonumber \end{eqnarray}
\begin{eqnarray}
   && \int d^3 x  D_{1-}D_{1+} \nonumber \\ && \times \left[   \frac{\delta  \Gamma^{1+ 2+ 0}}{\delta \Gamma_{1a}^* }\frac{\delta  \Gamma^{1+ 2+ 0}}{\delta \Gamma_{1a} } + 
\frac{\delta  \Gamma^{1+ 2+ 0}}{\delta c^*_1 }\frac{\delta  \Gamma^{1+ 2+ 0}}{\delta c_1 } + 
b_1 \frac{\delta  \Gamma^{1+ 2+ 0}}{\delta \bar c_1 }\right]_{\theta_1 =0} 
   \nonumber \\ &&  +\int d^3 x   D_{2-}D_{2+} \nonumber \\ && \times  \left[   \frac{\delta  \Gamma^{1+ 2+ 0}}{\delta \Gamma_{2a}^* }\frac{\delta  \Gamma^{1+ 2+ 0}}{\delta \Gamma_{2a} } + 
\frac{\delta  \Gamma^{1+ 2+ 0}}{\delta c^*_2 }\frac{\delta  \Gamma^{1+ 2+ 0}}{\delta c_2 } + 
b_2 \frac{\delta  \Gamma^{1+ 2+ 0}}{\delta \bar c_2 }\right]_{\theta_2 =0} \nonumber \\ &=&0
 ,
\nonumber \end{eqnarray}
\begin{eqnarray}
  &&
   \int d^3 x  D_{1-}D_{1+} \nonumber \\ && \times  \left[   \frac{\delta  \Gamma^{1+ 2+ 0}}{\delta \Gamma_{1a}^* }\frac{\delta  \Gamma^{1+ 2+ 0}}{\delta \Gamma_{1a} } + 
\frac{\delta  \Gamma^{1+ 2+ 0}}{\delta c^*_1 }\frac{\delta  \Gamma^{1+ 2+ 0}}{\delta c_1 } + 
b_1 \frac{\delta  \Gamma^{1+ 2+ 0}}{\delta \bar c_1 }\right]_{\theta_1 =0}  \nonumber \\
   && +\int d^3 x   D_{2+} D_{2-} \nonumber \\ && \times \left[  \frac{\delta  \Gamma^{1+ 2+ 0}}{\delta \Gamma_{2a}^* }\frac{\delta  \Gamma^{1+ 2+ 0}}{\delta \Gamma_{2a} } + 
\frac{\delta  \Gamma^{1+ 2+ 0}}{\delta c^*_2 }\frac{\delta  \Gamma^{1+ 2+ 0}}{\delta c_2 } + 
b_2 \frac{\delta  \Gamma^{1+ 2+ 0}}{\delta \bar c_2 }\right]_{\theta_2 =0} \nonumber \\ &=&0
 ,
\nonumber \end{eqnarray}
\begin{eqnarray}  
  &&\int d^3 x   D_{1+} D_{1-} \nonumber \\ && \times  \left[ \frac{\delta  \Gamma^{1+ 2+ 0}}{\delta \Gamma_{1a}^* }\frac{\delta  \Gamma^{1+ 2+ 0}}{\delta \Gamma_{1a} } + 
\frac{\delta  \Gamma^{1+ 2+ 0}}{\delta c^*_1 }\frac{\delta  \Gamma^{1+ 2+ 0}}{\delta c_1 } + 
b_1 \frac{\delta  \Gamma^{1+ 2+ 0}}{\delta \bar c_1 }\right]_{\theta_1 =0}  \nonumber \\ &&  
  + \int d^3 x  D_{2-}D_{2+} \nonumber \\ && \times  \left[   \frac{\delta  \Gamma^{1+ 2+ 0}}{\delta \Gamma_{2a}^* }\frac{\delta  \Gamma^{1+ 2+ 0}}{\delta \Gamma_{2a} } + 
\frac{\delta  \Gamma^{1+ 2+ 0}}{\delta c^*_2 }\frac{\delta  \Gamma^{1+ 2+ 0}}{\delta c_2 } + 
b_2 \frac{\delta  \Gamma^{1+ 2+ 0}}{\delta \bar c_2 }\right]_{\theta_2 =0} \nonumber \\ &=&0.
 \end{eqnarray}
It may be noted that it is possible to obtain higher order Slavnov-Taylor Identity for such theories. In fact, this procedure can be used 
to obtain Slavnov-Taylor Identity  for any gauge theory in presence of a boundary. This identity can be used to relate different correlation  
functions to each other. Thus, they can be used to analyzed scattering processes in this theory. It is important to note that this identity 
preserves only half of the supersymmetry of the original theory.

 \section{Conclusion}
 
In this paper, we analysed a three dimensional  supersymmetric theory   
with  $\mathcal{N} = 2$ supersymmetry. 
Even though the BRST symmetry has been analysed for a Yang-Mills theory with a boundary in $\mathcal{N} = 1$ superspace \cite{9}, in 
this paper, we analyse the BRST symmetry for a Yang-Mills theory with a  boundary in $\mathcal{N} = 2$ superspace. Furthermore, we 
analyse the BRST symmetry. 
The effective Lagrangian   was obtained by  by 
the sum of the gauge fixing term and the ghost term with the original  classical  
Lagrangian.
It was demonstrated that even though 
the supersymmetry of the effective Lagrangian  was  broken 
by the presence of the boundaries, it was possible to 
preserve half the supersymmetry of this effective Lagrangian  
This was done 
by adding new boundary terms to the original bulk effective Lagrangian.
The supersymmetric  variation of the original bulk effective Lagrangian was
exactly canceled by the the supersymmetric variation of this new boundary term. Thus, it was possible to retain 
the half of the supersymmetry of this original theory in presence of a boundary. We also obtain the  Slavnov-Taylor Identity  
for this theory. 
 
It may be noted that in  the Horava-Witten  theory,   
one of the low energy limits of the heterotic string theory can
be obtained from the  eleven dimensional supergravity in presence of a  boundary  
\cite{z12}-\cite{1z12ab}.
It has been  possible in this construction to obtain  a unification of 
gauge and gravitational couplings.
Motivated by the original Horava-Witten  theory, 
a five dimensional globally supersymmetric Yang-Mills theory coupled to a four dimensional hypermultiplet
on the boundary has already  been constructed  \cite{z21}. It would be interesting to use results of this 
paper to analyse such a system. It would also be interesting to analyse the BRST symmetry of such a system 
using both linear and non-linear gauges. Furthermore, the BRST symmetry and gauge fixing has been studied 
for perturbative quantum gravity \cite{pert0}-\cite{pert1}. It is possible to generalize this work to supergravity solutions, and 
analyse the supersymmetry of such supergravity solutions, when there is a boundary. In fact, the supergravity solutions 
with a boundary term have been studied, and this was done using a similar off-shell formalism \cite{supergrav}. It would be interesting to 
analyse the BRST symmetry for such supergravity theories with a boundary term.


\begin{thebibliography}{99}

 \bibitem{a} D. V. Belyaev, JHEP. 0601, 046 (2006)
 \bibitem{b} D. V. Belyaev, JHEP. 0601, 047 (2006)
 \bibitem{c}P. van Nieuwenhuizen and D. V. Vassilevich, Class. Quant. Grav. 22, 5029 (2005)
 \bibitem{d}U. Lindstrom, M. Rocek and P. van Nieuwenhuizen, Nucl. Phys. B 662, 147 (2003)
 \bibitem{e} P. Di Vecchia, B. Durhuus, P. Olesen and J. L. Petersen, Nucl. Phys. B 207, 77 (1982)

 \bibitem{f}P. Di Vecchia, B. Durhuus, P. Olesen and J. L. Petersen, Nucl. Phys. B 217, 395 (1983)
 \bibitem{g}Y. Igarashi, Phys. Rev. D 30, 1812 (1984)

\bibitem{1} D.  V. Belyaev and P. van Nieuwenhuizen,  JHEP.   0804, 008 (2008)

\bibitem{4} D. S. Berman and D. C Thompson, Nucl. Phys. B820, 503 (2009)
\bibitem{5}M. Faizal and D. J. Smith, Phys. Rev. D85, 105007 (2012)
\bibitem{5a}M. Faizal, Mod. Phys. Lett. A29,  1450154  (2014)
\bibitem{6}M. Faizal, JHEP 1204, 017 (2012)
 

\bibitem{8}M. Faizal and D. J. Smith, Phys. Rev. D87, 025019  (2013)
\bibitem{9}M. Faizal,   Int. J. Theor. Phys. 52, 392  (2013)

\bibitem{brst} C. Becchi, A. Rouet and R. Stora,  Annals. Phys. 98, 287  (1976) 
\bibitem{brst1} I. V. Tyutin,  Lebedev. preprint. fian. 39 (1975) 
\bibitem{nlbrst}D. Dudal, H. Verschelde, V. E. R. Lemes, M. S. Sarandy, S. P. Sorella and M. Picariello, 
Ann. Phys. 308, 62 (2003)
\bibitem{dj1} R. Delbourgo and P. D. Jarvis, J. Phys. A  Math. Gen. 15, 6 11 (1982)
\bibitem{dj} M.  Faizal and S.  Upadhyay,  Phys. Lett. B 736,  288 (2014)
\bibitem{dj2}  L. Baulieu and J. Thierry-Mieg, Nucl. Phys. B 197, 477 (1982)

\bibitem{nlbrst1}D. Dudal, V.E.R. Lemes, M. Picariello, M.S. Sarandy, S.P. Sorella and H. Verschelde, 
JHEP. 0212, 008 (2002)
\bibitem{mir} M. Faizal, Phys. Rev. D 84, 106011 (2011) 
\bibitem{0mir1}M. Faizal, Comm. Theor. Phys.  57, 637 (2012)  
\bibitem{0mir2}M. Faizal and D. J. Smith,  Phys. Rev. D85,  105007  (2012)   
\bibitem{0mir4}M. Faizal, Mod. Phys. Lett. A27, 1250147  (2012) 
\bibitem{sudd} S. Upadhyay and D. Das, Phys. Lett. B 733, 63 (2014)  
\bibitem{kon} K. I. Kondo, arXiv: 0103141 
\bibitem{z1} D. Dudal, H. Verschelde, V. E. R. Lemes, M. S. Sarandy, S. P. Sorella, M. Picariello, A. Vicini and J. A. Gracey, 
JHEP 0306, 003  (2003)  
\bibitem{z2} D. Dudal and H.  Verschelde,  J. Phys. A36, 8507 (2003)  
\bibitem{z3} V. E. R. Lemes, M. S. Sarandy and S. P. Sorella, Ann. Phys. 308, 1 (2003) 
\bibitem{z4}A. R. Fazio, V. E. R. Lemes, M. Picariello, M. S. Sarandy,  and S. P. Sorella, Mod. Phys. 
Lett. A18, 711 (2003) 
\bibitem{sch} M. Schaden,   arXiv: 9909011 
\bibitem{kon1} K.-I. Kondo and T. Shinohara,   Phys. Lett. B 491, 263 
(2000) 
\bibitem{hooft} G. 't Hooft,   Nucl.Phys. B 190, 455  (1981) 
\bibitem{nambu} Y. Nambu,   Phys. Rev. D 10, 4262 (1974) 
 \bibitem{mandal} S. Mandelstam,  Phys. Report 23, 245 (1976) 
\bibitem{poly} A. M. Polyakov,   Nucl. Phys. B 120, 429  (1977) 
\bibitem{kon2}K. I. Kondo,   Phys. Rev. D 58, 085013 (1998) 
\bibitem{kon3}K. I. Kondo,   Phys. Rev. D 58, 105016 (1998) 
\bibitem{kon4}K. I. Kondo,   Phys. Lett. B 455, 251  (1999) 
\bibitem{n0}L. Baulieu and S. P. Sorella, Phys. Lett. B 671, 481 (2009) 
\bibitem{n1}L. Baulieu, M. A. L. Capri, A. J. Gomez, V. E. R. Lemes, R. F. Sobreiro and  S. P. Sorella, 
Eur. Phys. J. C66, 451 (2010) 
\bibitem{n2}D. Dudal, S.P. Sorella, N. Vandersickel and H. Verschelde, Phys. Rev. D 79, 121701 (2009) 
\bibitem{n4} P. M. Lavrov, O. V. Radchenko and A. A. Reshetnyak,  Mod. Phys. Lett. A. 27,   1250067 (2012) 
\bibitem{gate} S. J. Gates Jr , M. T. Grisaru, M. Rocek and W. Siegel, 
  Front. Phys. 58,  1  (1983)
\bibitem{nonc} D. N. Blaschke, H. Grosse and J. C. Wallet, JHEP 1306 038 (2013)
 \bibitem{z12} P. Horava and E. Witten,  Nucl. Phys. B 475, 94 (1996)
 \bibitem{z5ab}E. Witten, Nucl. Phys. B 471, 135 (1996)
 \bibitem{z41ab}Horava, Phys. Rev. D 54, 7561 (1996) 
 \bibitem{1z12ab} P. Horava and E. Witten, Nucl. Phys. B 460, 506 (1996)
 \bibitem{z21}E. A. Mirabelli and M. E. Peskin,   Phys. Rev. D 58, 065002 (1998)
 \bibitem{pert0}M. Faizal, Found. Phys. 41, 270 (2011) 
 \bibitem{pe} G. Esposito, G. Fucci, A. Y. Kamenshchik and K. Kirsten, Class. Quant. Grav. 22, 957 (2005) 
 \bibitem{ep}  G. Esposito, A. Yu. Kamenshchik, I. V. Mishakov and  G. Pollifrone, Phys. Rev. D 52, 3457 (1995) 
 \bibitem{2pert}M. Faizal, Class. Quant. Grav. 29, 035007 (2012) 
 \bibitem{4pert}M. Faizal, Mod. Phys. Lett. A 28, 1350034 (2013) 
 \bibitem{pert1}M. Faizal,  J. Phys. A 44, 402001 (2011)
 \bibitem{supergrav} D. V. Belyaev and  P. van Nieuwenhuizen,  	JHEP 0809, 069 (2008)
 \end{thebibliography}
\end{document}